\documentclass[12pt,twoside]{article}
\usepackage[mathscr]{eucal}
\usepackage{amsmath,amsfonts,amssymb,amsthm,mathabx,empheq}
\usepackage{times}
\usepackage{pdfsync}
\usepackage{cite}
\usepackage{url}
\usepackage{hyperref}
\usepackage{tensor}
\usepackage{color}
\usepackage{multicol}
\usepackage{bbold}

\advance\textheight\topskip
\allowdisplaybreaks[1]

\newcommand\td{\text{d}}
\newcommand\cO{{\cal O}}
\newcommand{\p}{\partial}

\newcommand{\be}{\begin{equation}}
\newcommand{\ee}{\end{equation}}
\newcommand{\bea}{\begin{eqnarray}}
\newcommand{\eea}{\end{eqnarray}}

\def\bz{\bar z}

\def\pb{\bar \partial}
\def\bp{\bar \partial}

\def\bm{\bar{m}}
\def\bP{\bar P}

\newcommand*\xbar[1]{%
  \hbox{%
    \vbox{%
      \hrule height 0.5pt 
      \kern0.3ex
      \hbox{%
        \kern-0.0em
        \ensuremath{#1}%
        \kern-0.0em
      }%
    }%
  }%
}

\allowdisplaybreaks[1]

\DeclareFontFamily{OT1}{rsfs}{} \DeclareFontShape{OT1}{rsfs}{m}{n}{
<-7> rsfs5 <7-10> rsfs7 <10-> rsfs10}{}
\DeclareMathAlphabet{\mycal}{OT1}{rsfs}{m}{n}

\hypersetup{
colorlinks=true,
linktoc=page,    
linkcolor=blue,
citecolor=blue,
urlcolor=blue,
colorlinks=true,
}

\begin{document}

\title{Asymptotic Weyl double copy in Newman-Penrose formalism}

\author{Pujian Mao and Weicheng Zhao}

\date{}

\def\mytitle{Asymptotic Weyl double copy in Newman-Penrose formalism}

\addtolength{\headsep}{4pt}

\begin{centering}

  \vspace{1cm}

  \textbf{\large{\mytitle}}

  \vspace{1.5cm}

  {\large Pujian Mao and Weicheng Zhao }

\vspace{0.5cm}

\begin{minipage}{.9\textwidth}\small \it  \begin{center}
     Center for Joint Quantum Studies and Department of Physics,\\
     School of Science, Tianjin University, 135 Yaguan Road, Tianjin 300350, China
 \end{center}
\end{minipage}

\vspace{0.3cm}

\end{centering}

\begin{center}
Emails: pjmao@tju.edu.cn,\, zhaoweichengok@tju.edu.cn
\end{center}

\begin{center}
\begin{minipage}{.9\textwidth}
\textsc{Abstract}: In this paper, we provide a self-contained investigation of the Weyl double copy in the Newman-Penrose formalism. We examine the Weyl double copy constraints for the general asymptotically flat solution in the Newman-Unti gauge. We find that two transparent solutions of the asymptotic Weyl double copy constraints lead to truncated solutions for both linearized and Einstein gravity theory where the solutions are in the manifest form of Petrov type N or type D in the Newman-Unti gauge. 

\end{minipage}
\end{center}

\thispagestyle{empty}


\section{Introduction}

Weyl double copy (WDC) \cite{Luna:2018dpt} is a fascinating realization of the classical double copy relation \cite{Monteiro:2014cda} which in general connects solutions in gravitational theories with gauge theories. The WDC formula involves only
curvatures, namely the Weyl tensor and the Maxwell field strength tensor, rather than the gauge fields. Hence, it is adorable from its gauge invariant nature and has been an active research topic \cite{Alawadhi:2020jrv,Godazgar:2020zbv,White:2020sfn,Monteiro:2020plf,Chacon:2021wbr,Godazgar:2021iae,Adamo:2021dfg,Easson:2021asd,Chacon:2021lox,Han:2022ubu,Han:2022mze,Luna:2022dxo,Easson:2022zoh,Easson:2023dbk,Alkac:2023glx}. The WDC relations are best expressed in spinor language as
\begin{equation}\label{WDC}
    \Psi_{ABCD}=\frac{3c}{S}\Phi_{(AB}\Phi_{CD)}.
\end{equation}
The formula of the WDC is very elegant, the applications are somewhat restrictive. On the one hand, by construction, the WDC formula is especially for Petrov type D \cite{Luna:2018dpt} and Petrov type N \cite{Godazgar:2020zbv} solutions. On the other hand, the spinorial description restricts its extensions to theories in dimensions other than 4 or to other types of gravitational theory. In this paper, we provide an alternative probe of the WDC by exploring its constraints on the solutions of gravity and Maxwell theory in the Newman-Penrose (NP) formalism \cite{Newman:1961qr}.

The investigation in this work is mainly inspired by the asymptotic WDC \cite{Godazgar:2021iae}. In the NP formalism, there is a well known solution space in the form of a series expansion in the inverse radial coordinate in the Newman-Unti (NU) gauge \cite{Newman:1962cia}. Solutions of the Maxwell theory in the NP formalism are also known in the same type of series expansion \cite{Janis:1965tx,Newman:1968uj}. The main concern of this work is to examine the consequence of the WDC relations on those two solution spaces asymptotically. We first present a direct verification in the NP formalism that the WDC formula is invariant under the local Lorentz transformations and the solution satisfying the WDC formula must be of either Petrov type D or type N. Those properties are transparent in the spinor language. Nevertheless, they are still meaningful for a self-contained investigation in the NP formalism. We then list the asymptotic WDC constraints on the solution space of Maxwell theory and linearized gravity. There are several reasons for linearization for the gravitational theory. The solution space of the linearized theory is much simpler and tractable which will manifest the WDC constraints. We linearize the theory on Minkowski background in the present work. Nonetheless, it is straightforward to perform a similar investigation on a curved background which extends the original scope of the WDC. The classical double copy is originated from amplitude double copye relation \cite{Kawai:1985xq,Bern:2008qj,Bern:2010ue} where gravity amplitude is defined perturbatively on Minkowski background. So the linearization at the classical level may better approach the quantum feature. There are four equations from the asymptotic WDC relations constraining the time evolution of the scalar field in the WDC formula. A generic solution to those constraints is a very challenging problem. Nevertheless, there are some interesting special solutions. We check two special cases. Interestingly, those two solutions lead to significant simplification of the solution space and result in truncated solutions in a manifest form of type N or type D in the NU gauge. Remarkably, those two simplifications with truncation sustain at the non-linear level. When we choose the asymptotic WDC conditions as initial data, the solutions of the NP system that satisfy the WDC relations in the NU gauge truncate and are in the exact form of type N or type D.

The organization of this paper is as follows. In the next section, we examine the properties of the WDC in the NP formalism. In Section \ref{solution}, we list the solution space of Maxwell theory and linearized gravity on Minkowski background. In Section \ref{WDCrelation}, we present the asymptotic WDC constraints on the solution space and find that two solutions of the WDC constraints lead to truncation of the solution space. In Section \ref{full}, we show that the same type of truncation also happens for full Einstein gravity. We conclude in the last section. There is one Appendix that details our conventions for the NP formalism.


\section{Weyl double copy in NP formalism}

In the NP formalism,\footnote{See Appendix \ref{NP} for our conventions.} after choosing an appropriate system of null tetrads, the spinor form of the WDC can be translated into the connections of the Weyl scalars and the Maxwell scalars as follows \cite{Godazgar:2021iae}:
\begin{equation}\label{wdcnp}
    \begin{split}
        &\Psi_4=\frac{3c}{S}\phi_2\phi_2,  \phantom{10000} \Psi_3=\frac{3c}{S}\phi_1\phi_2, \phantom{10000} \Psi_2=\frac{c}{S}(\phi_0\phi_2+2\phi_1\phi_1),\\
        &\phantom{100000000}\Psi_1=\frac{3c}{S}\phi_0\phi_1,
        \phantom{10000} \Psi_0=\frac{3c}{S}\phi_0\phi_0.  
    \end{split}
\end{equation}
Since the null vector $l$ can be transformed to any null direction through a single second class null rotation and a single first class null rotation, we can verify that the WDC relations are invariant under such gauge transformations. Under a combined first and second classes of null rotations, the Weyl scalars and Maxwell scalars transform into:
\begin{equation}
    \begin{split}
        &\Psi_0'\rightarrow \Psi_0+4b\Psi_1+6b^2\Psi_2+4b^3\Psi_3+b^4\Psi_4,\\  
        &\Psi_1'\rightarrow \bar{a}\Psi_0+(1+4\bar{a}b)\Psi_1+3b(1+2\bar{a}b)\Psi_2+b^2(3+4\bar{a}b)\Psi_3+b^3(1+\bar{a}b)\Psi_4,\\
        &\Psi_2'\rightarrow \bar{a}^2\Psi_0+2\bar{a}(1+2\bar{a}b)\Psi_1+(1+6\bar{a}b+6\bar{a}^2 b^2)\Psi_2\\
        &\hspace{5cm}+2b(1+3\bar{a}b+2\bar{a}^2 b^2)\Psi_3+b^2(1+{a}b)^2\Psi_4,\\
        &\Psi_3'\rightarrow    \bar{a}^3\Psi_0+\bar{a}^2(3+4\bar{a}b)\Psi_1+3\bar{a}(1+3\bar{a}b+2\bar{a}^2b^2)\Psi_2\\
        &\hspace{4cm}+(1+6\bar{a}b+9\bar{a}^2 b^2+4\bar{a}^3 b^3)\Psi_3+b(1+\bar{a}b)^3\Psi_4,\\ &\Psi_4'\rightarrow \bar{a}^4 \Psi_0+4\bar{a}^3(1+\bar{a}b)\Psi_1+6\bar{a}^2(1+\bar{a}b)^2\Psi_2+4\bar{a}(1+\bar{a}b)^3\Psi_3+(1+\bar{a}b)^4\Psi_4,\\
        &\phi_2'\rightarrow \bar{a}^2\phi_0+2\bar{a}(1+\bar{a}b)\phi_1+(1+\bar{a}b)^2\phi_2,\\
        &\phi_1'\rightarrow \bar{a}\phi_0+(2\bar{a}b+1)\phi_1+b(1+\bar{a}b)\phi_2,\\
        &\phi_0'\rightarrow  \phi_0+2b\phi_1+b^2\phi_2,
        \end{split}
\end{equation}
where $a$ and $b$ are complex functions that characterize first and second class null rotations respectively. It is straightforward to verify that, once the WDC relations are satisfied in the original frame, they sustain in the new frame,
\begin{equation}
    \begin{split}
                &\Psi_4'=\frac{3c}{S}\phi_2'\phi_2',  \phantom{10000} \Psi_3'=\frac{3c}{S}\phi_1'\phi_2', \phantom{10000} \Psi_2'=\frac{c}{S}(\phi_0'\phi_2'+2\phi_1'\phi_1'),\\
        &\phantom{100000000}\Psi_1'=\frac{3c}{S}\phi_0'\phi_1',
        \phantom{10000} \Psi_0'=\frac{3c}{S}\phi_0'\phi_0' . 
    \end{split}
\end{equation}
Assuming that $S$ is invariant under the null rotations, this is a direct confirmation in the NP formalism that the WDC formula is invariant under local Lorentz transformations. Then, it is easy to verify that any gravitational solution that satisfies the WDC formula must necessarily be of either Petrov type D or Petrov type N. This is because the WDC formula requires that the Weyl scalars can be expressed as products of Maxwell scalars. Under the second class null rotation, the Maxwell scalars transform as
\begin{equation}
    \begin{split}
        &\phi_2'=\phi_2,\\
        &\phi_1'=\phi_1+b\phi_2,\\
        &\phi_0'=\phi_0+2b\phi_1+b^2\phi_2.
    \end{split}
\end{equation}
For the condition $\phi_0+2b\phi_1+b^2\phi_2= 0$, if $b$ has a double root, we can use a second class null rotation to make $\phi_0' = \phi_1' = 0$. Accordingly, the gravitational field constructed from Maxwell field from \eqref{wdcnp} in the new frame system is manifest of Type N. If $\phi_0+2b\phi_1+b^2\phi_2= 0$ has two distinct roots, then we can always use another first class null rotation to make $\phi_0'' = \phi_2'' = 0$. Accordingly, the gravitational solution is of type D.


\section{Solution space}
\label{solution}

In this section, we will review the solution space for Maxwell theory and linearized gravity with adaption to our conventions. Relevant results were presented in \cite{Janis:1965tx,Newman:1968uj}, also in \cite{Conde:2016csj,Conde:2016rom}. For computational simplicity, we use flat null coordinates $x^\mu=(u,r,z,\bz)$ \cite{He:2019jjk} where the celestial sphere at null infinity is mapped to a 2d plane \cite{Barnich:2016lyg,Compere:2016jwb,Compere:2018ylh,Barnich:2021dta}. The line element of the Minkowski spacetime in this coordinates system is
\be\label{flat}
\td s^2=-2 \td u \td r + 2r^2 \td z \td \bz .
\ee
The only non-vanishing NP variables are $\rho=-\frac1r$ and $L^{\bz}=\frac{1}{r}$. Correspondingly, the directional derivatives are given by
\begin{align}
D=\frac{\p}{\p r},\quad \Delta=\frac{\p}{\p u},\quad \delta=\frac1r \frac{\p}{\p \bz}.
\end{align}
We define $\p=\p_{z}$ and $\bar\p=\p_{\bz}$ for notational brevity.

\subsection{Maxwell theory}

Maxwell's equations in the flat null coordinates are organized as
\be
\p_r (r^2 \phi_1) =r\p \phi_0,\quad \p_r (r \phi_2) =\p \phi_1,\quad \p_u \phi_0=\frac1r \bp \phi_1,\quad \p_u \phi_1=\frac1r \bp \phi_2. \label{M}
\ee
Suppose that $\phi_0$ is given as initial data in series expansion,
\be
\phi_0=\frac{\phi_0^0}{r^3}+\sum_{i=1}^{\infty}\frac{\phi_0^i}{r^{i+3}}.
\ee
Then $\phi_1$ and $\phi_2$ can be solved as
\begin{align}
\phi_1=\frac{\phi_1^0}{r^2}-&\frac{\p \phi_0^0}{r^3}-\sum_{i=1}^\infty\frac{1}{(i+1)} \frac{\p \phi_0^i}{r^{i+3}},\\
\phi_2=\frac{\phi_2^0}{r}-\frac{\p\phi_1^0}{r^2}+&\frac{\p^2 \phi_0^0}{2r^3}+\sum_{i=1}^\infty\frac{1}{(i+1)(i+2)} \frac{\p^2 \phi_0^i}{r^{i+3}}.
\end{align}
The evolution of the initial data is controlled by
\be
\begin{split}
 &\p_u \phi_1^0=\bp \phi_2^0,\\
 &\p_u \phi_0^0=\bp \phi_1^0,\\
 &\p_u \phi_0^i=-\frac{1}{i}\bp\p\phi_0^{i-1},\quad i \geq 1.
\end{split}
\ee

\subsection{Linearized gravity theory}

Bianchi identities after the linearization are given by\footnote{We use $\psi_i$ to denote the Weyl scalars in the linearized theory.}
\be
\begin{split}
&\p_r (r^4 \psi_1)=r^3 \p \psi_0,\quad \p_r (r^3 \psi_2)=r^2 \p \psi_1,\quad \p_r (r^2 \psi_3)=r \p \psi_2,\quad \p_r (r \psi_4)= \p \psi_3,\\ 
&\p_u \psi_0 = \frac1r \pb \psi_1,\quad \p_u \psi_1 = \frac1r \pb \psi_2,\quad \p_u \psi_2 = \frac1r \pb \psi_3,\quad \p_u \psi_3 = \frac1r \pb \psi_4. 
\end{split}
\ee
Suppose that $\psi_0$ is given as initial data in series expansion,
\be
\psi_0=\frac{\psi_0^0}{r^5}+\sum_{i=1}^{\infty}\frac{\psi_0^i}{r^{i+5}}.
\ee
Then $\psi_1$-$\psi_4$ can be solved as
\be\label{linearized}
\begin{split}
\psi_1=\frac{\psi_1^0}{r^4}-&\frac{\p \psi_0^0}{r^5}-\sum_{i=1}^\infty\frac{1}{(i+1)} \frac{\p \psi_0^i}{r^{i+5}},\\
\psi_2=\frac{\psi_2^0}{r^3}-\frac{\p\psi_1^0}{r^4}+&\frac{\p^2 \psi_0^0}{2r^5}+\sum_{i=1}^\infty\frac{1}{(i+1)(i+2)} \frac{\p^2 \psi_0^i}{r^{i+5}},\\
\psi_3=\frac{\psi_3^0}{r^2}-\frac{\p\psi_2^0}{r^3}+\frac{\p^2 \psi_1^0}{2r^4}-&\frac{\p^3 \psi_0^0}{6r^5}-\sum_{i=1}^\infty\frac{1}{(i+1)(i+2)(i+3)} \frac{\p^3 \psi_0^i}{r^{i+5}},\\
\psi_4=\frac{\psi_4^0}{r}-\frac{\p\psi_3^0}{r^2}+ \frac{\p^2\psi_2^0}{2r^3}- \frac{\p^3\psi_1^0}{6r^4}+&\frac{\p^4 \psi_0^0}{24r^5}+\sum_{i=0}^\infty\frac{1}{(i+1)(i+2)(i+3)(i+4)} \frac{\p^4 \psi_0^i}{r^{i+5}}.
\end{split}
\ee
The evolution of the initial data is controlled by
\be\label{evolutionW}
\begin{split}
 &\p_u \psi_3^0=\bp \psi_4^0,\\
 &\p_u \psi_2^0=\bp \psi_3^0,\\
 &\p_u \psi_1^0=\bp \psi_2^0,\\
 &\p_u \psi_0^0=\bp \psi_1^0,\\
 &\p_u \psi_0^i=-\frac{1}{i}\bp\p\psi_0^{i-1},\quad i \geq 1.
\end{split}
\ee


\section{WDC constraints and linearized gravity}
\label{WDCrelation}

In this section, we will examine the asymptotic WDC constraints on the solution of Maxwell and linearized gravitational theory asymptotically. For our purpose, it is convenient to have the scalar field in the WDC formula on the numerator,
\begin{equation}\label{awdc}
    \begin{split}
        \psi_4=F\phi_2\phi_2,  \quad &\psi_3=F\phi_1\phi_2, \quad \psi_2= \frac13 F(\phi_0\phi_2+2\phi_1\phi_1),\\
        \quad&\psi_1=F\phi_0\phi_1,
        \quad \psi_0=F\phi_0\phi_0,  
    \end{split}
\end{equation}
where $F$ can be expanded in inverse powers of $r$ as
\be
F=F_0 r + \sum_{i=1}^\infty\frac{F_i}{r^{i-1}}.
\ee
The asymptotic WDC was proposed in \cite{Godazgar:2021iae} by requiring that the leading orders of all Weyl scalars and Maxwell scalars satisfy the WDC formula. If we assume that the Maxwell scalars already satisfy the Maxwell's equations and construct the Weyl scalars via the asymptotic WDC, then the Bianchi identities in \eqref{evolutionW} yield the following constraints  at the leading orders,
\begin{equation}
    \begin{split}
        &\phi_1^0\p_u(F_0 \phi_2^0)=\phi_2^0 \bar{\p}(F_0 \phi_2^0),\\
        &2(\phi_1^0)^2 \p_u F_0 +\phi_0^0 \p_u(F_0 \phi_2^0)=-F_0\bar{\p}(\phi_1^0\phi_2^0)+3\phi_2^0\bar{\p}(F_0\phi_1^0),\\
        &3\p_u F_0\phi_1^0\phi_0^0=-2F_0 \phi_0^0\bar{\p}\phi_2^0+F_0 \phi_2^0\bar{\p}\phi_0^0+\phi_0^0 \phi_2^0 \bar{\p}F_0 +F_0 \phi_1^0 \bar{\p}\phi_1^0+2(\phi_1^0)^2\bar{\p}F_0,\\
        &\phi_0^0 \p_u (\phi_0^0 F_0)=\phi_1^0\bar{\p}(F_0 \phi_0^0 ).
    \end{split}
\end{equation}
The above four equations can all be seen as evolution equations for $F_0$. 
Since the system is way more overdetermined, direct solutions are challenging. Nevertheless, we will consider some special cases. There are two transparent solutions, $\phi_1^0 = \phi_0^0 = \bp F_0= 0,\, \phi_2^0\neq0$ and $\phi_2^0 = \phi_0^0 = \p_u F_0=\pb F_0= 0,\, \phi_1^0\neq0$. Surprisingly, as we will demonstrate below, those two simple solutions lead to truncation of the solution space when combined with the WDC relations. For the first option, the Maxwell fields are reduced to
\begin{equation}
    \begin{split}
        &\phi_0=\sum_{i=1}^{\infty}\frac{\phi_0^i}{r^{i+3}},\\
        &\phi_1=-\sum_{i=1}^\infty\frac{1}{(i+1)} \frac{\p \phi_0^i}{r^{i+3}},\\
        &\phi_2=\frac{\phi_2^0(u,z)}{r}+\sum_{i=1}^\infty\frac{1}{(i+1)(i+2)} \frac{\p^2 \phi_0^i}{r^{i+3}}.
    \end{split}
\end{equation}
Then, WDC relations yield that
\begin{equation}\label{awdcn}
    \begin{split}
        &\psi_4=\frac{F_0 \phi_2^0\phi_2^0}{r}+O(r^{-4}),\quad
        \psi_3=-\frac{F_0 \phi_2^0\p\phi_0^1}{2r^4}+O(r^{-5}),\quad
        \psi_2=\frac{F_0\phi_2^0\phi_0^1}{r^4}+O(r^{-5}),\\  
        &\phantom{100000000000}\psi_1=-\frac{F_0\phi_0^1\p\phi_0^1}{r^7}+O(r^{-8}),\quad
        \psi_0=\frac{F_0\phi_0^1\phi_0^1}{r^7}+O(r^{-8}).
    \end{split}
\end{equation}
Clearly, the WDC construction set that $\psi_3^0=\psi_2^0=\psi_1^0=\psi_0^0=\psi_0^1=0$. Hence, $\psi_0$ begins at order $O(r^{-7})$. As evident from the solution space of the linearized gravity \eqref{linearized}, those conditions lead to that all the Weyl scalars should begin at order $O(r^{-7})$ except for $\psi_4$. Consequently, the expression of $\psi_2$ in \eqref{awdcn} indicates that $\phi_0^1=0$. In fact, since $\phi_2^0\neq0$, the leading term of $\psi_2$ is $\frac{F_0\phi_2^0\phi_0^i}{r^{i+3}}$, and the leading term of $\psi_0$ is $\frac{F_0\phi_0^i \phi_0^i}{r^{2i+5}}$ from the asymptotic WDC formula \eqref{awdcn}, where $\phi_0^i$ is the leading non-vanishing order of $\phi_0$. However, in the solution space \eqref{linearized}, it is clear that $\psi_2$ and $\psi_0$ should begin at the same order when $\psi_2^0=0$. This yields that $\phi_0^i=0$ for any $i$, namely $\phi_0=0$. Hence, $\phi_1=0$. Thus, the electromagnetic solution is type N and $l$ is the principle null direction. The WDC formula implies that any asymptotically type N electromagnetic solution is manifestly type N. Finally, the full solutions fulfilling the WDC relations are
\begin{equation}
\begin{split}
       &\phi_0=\phi_1=0, \quad\phi_2=\frac{\phi_2^0(u,z)}{r},\\ 
       &\psi_0=\psi_1=\psi_2=\psi_3=0,\quad \psi_4=\frac{\psi_4^0(u,z)}{r},\\
       &F=F_0 r,\quad F_0=\frac{\psi_4^0}{\phi_2^0\phi_2^0}.
\end{split}
\end{equation}
Going back to the original WDC relations in \eqref{wdcnp}, we obtain $S=\frac{3c\phi_2^0 \phi_2^0}{\psi_4^0 r}$, which is just the news function of a scalar field \cite{Bondi:1962px}. It is obvious that $S$ satisfies the equation of motion of a scalar field on Minkowski spacetime.

For the second option, the Maxwell fields are reduced to
\begin{equation}
\begin{split}
    &\phi_0=\frac{\phi_0^1(z,\bar{z})}{r^4}+O(r^{-5}),\\
    &\phi_1=\frac{\phi_1^0(z)}{r^2}-\frac{1}{2}\frac{\p\phi_0^1}{r^4}+O(r^{-5}),\\
    &\phi_2=-\frac{\p\phi_1^0(z)}{r^2}+\frac{1}{6}\frac{\p^2\phi_0^1}{r^4}+O(r^{-5}).
\end{split}
\end{equation}
The WDC relations imply that
\begin{equation}
    \begin{split}
        &\psi_0=\frac{F_0\phi_0^1\phi_0^1}{r^7}+O(r^{-8}),\quad \psi_1=\frac{F_0\phi_0^1\phi_1^0}{r^5}+O(r^{-6}),\quad \psi_2=\frac{2F_0\phi_1^0\phi_1^0}{3r^3}+O(r^{-4}),\\
        &\phantom{10000000000}\psi_3=-\frac{F_0\phi_1^0\p\phi_1^0}{r^3}O(r^{-4}) ,\quad  \psi_4=\frac{F_0(\p\phi_1^0)^2}{r^3}+O(r^{-4}).
    \end{split}
\end{equation}
To be a solution of linearized gravity, one must set $F_0=F_0(z)$, and $\psi_4^0=\psi_3^0=\psi_1^0=\psi_0^0=\psi_0^1=0$.
Because of $\phi_1^0\neq0$ and $\phi_2^0=\phi_0^0=0$, the leading order of $\psi_1$ is $\frac{F_0\phi_1^0\phi_0^i}{r^{i+4}}$, while that of $\psi_0$ is $\frac{F_0\phi_0^i\phi_0^i}{r^{2i+5}}$. Similar to the case of the previous option, one must have that $\phi_0=0$. Then the solutions of the Maxwell theory are
\be
\phi_0=0,\quad \phi_1=\frac{\phi_1^0(z)}{r^2},\quad \phi_2=-\frac{\p\phi_1^0}{r^2}.
\ee
The gravitational fields constructed from the WDC relations are
\begin{equation}
\psi_0=\psi_1=0,\quad \psi_2=\frac{2}{3}\frac{F_0(\phi_1^0)^2}{r^3},\quad \psi_3=-\frac{F_0\phi_1^0\p\phi_1^0}{r^3},\quad \psi_4=\frac{F_0(\p\phi_1^0)^2}{r^3}.
\end{equation}
The solution of the linearized gravitational theory \eqref{linearized} finally determines that
\begin{equation}
    \phi_1^0=\frac{k_2}{(z+k_3)^2}, \quad F_0= \frac{k_1}{z+k_3},
\end{equation}
with $k_1,k_2,k_3$ being constants. The corresponding scalar field $S$ will satisfy the equation of motion. One can apply a first class of null rotation to have $\phi_2=0$. Then, the gravitational solution is in the manifest form of type D.


\section{Extension to Einstein gravity}
\label{full}

In the previous section, we have shown that two types of boundary conditions which are transparent solutions to the asymptotic WDC constraints lead to interesting truncated solution space for Maxwell and linear gravitational theory when combined with the WDC relations. In this section, we will prove that the same types of truncation also hold for the Einstein theory and Maxwell theory on the curved background spacetime as a probe.

\subsection{Type N solution}
In this subsection, we will demonstrate that, when choosing the boundary conditions $\phi_1^0=\phi_0^0=0$ and $\phi_2^0\neq0$ in asymptotically flat spacetime, the WDC relations \eqref{wdcnp} yield a truncated solution space which is manifestly of type N in the NU gauge. We first prove the following three lemmas from the WDC relations.

\noindent
\textbf{Lemma 1: If $\phi_0^0=\phi_1^0=0$, $\phi_2^0\neq0$, then $\phi_0^1=0$.}

\noindent
Proof: In this case, the leading terms of the Maxwell scalars on an asymptotically flat background spacetime are
\begin{equation}
    \phi_2=\frac{\phi_2^0}{r}+O(r^{-3}), \quad \phi_1=-\frac{\xbar{\eth}\phi_0^1}{2r^4}+O(r^{-5}),\quad \phi_0=\frac{\phi_0^1}{r^4}+O(r^{-5}).
\end{equation}
where the definitions of the $\eth$ and $\xbar\eth$ operators are presented below in \eqref{eth}. The WDC relations yield the Weyl scalars as
\begin{equation}
    \Psi_2=\frac{F_0 \phi_2^0 \phi_0^1}{r^4} +O(r^{-5}), \quad \Psi_1= -\frac{F_0 \phi_0^1\xbar\eth \phi_0^1}{2r^7}+O(r^{-8}),\quad \Psi_0=\frac{F_0(\phi_0^1)^2}{r^7}+O(r^{-8}).
\end{equation}
However, the two radial equations from the Bianchi identity \eqref{R14} and \eqref{R15}, with asymptotically flat fall-off conditions \cite{Newman:1962cia}, determine that $\Psi_2=\cO(r^{-7})$. This requires that $\phi_0^1=0$.

\noindent
\textbf{Lemma 2: If $\phi_1^0=0$ and $\phi_0=0$, then both the Maxwell fields and the gravitational fields are in the manifest form of Type N in the NU gauge.}

\noindent
Proof: According to the WDC relations, it is obvious from the assumptions that $\Psi_0=\Psi_1=0$. Therefore, the Goldberg-Sachs theorem yields that $\sigma=0$, hence $\rho=-\frac{1}{r}$. The Maxwell's equation \eqref{R18} yields that $\phi_1=0$. Consequently, both the Maxwell and the corresponding gravitational field are type N and $l$ is the principle null direction.

\noindent
\textbf{Lemma 3: If $\phi_1^0=\phi_0^0=0$, and $\phi_0^1=...=\phi_0^{i}=0$, then $\phi_1^1=...=\phi_1^i=0$ for $i \geq 1$.}

\noindent
Proof: We just need to apply the radial equation \eqref{R18} and the fact that the background spacetime is asymptotically flat. Lemma 3 can be proved easily.

Now we can prove that the WDC relations combined with the boundary condition $\phi_1^0=\phi_0^0=0$ require that the solutions are of type N and $l$ is the principle null direction. We know from the WDC relations that $\Psi_2=\frac{1}{3}F\phi_2\phi_0+\frac{2}{3}F(\phi_1)^2$, therefore the leading term of $\Psi_2$ could be either $\frac{F_0\phi_2^0\phi_0^i}{3r^{i+3}}$ or $\frac{2F_0(\phi_1^j)^2}{3r^{2j+3}}$, where $\phi_0^i$ and $\phi_1^j$ are the leading terms of $\phi_0$ and $\phi_1$, respectively. The Lemma 3 requires that the second term is always order-suppressed. Then the leading term of $\Psi_2$ is $\frac{F_0\phi_2^0\phi_0^i}{3r^{i+3}}$. While the WDC relation yields that the leading term of $\Psi_0$ is $\frac{F_0(\phi_0^i)^2}{r^{2i+5}}$. Thus the peeling-off property encoded in the radial equations \eqref{R14} and \eqref{R15} yields that $\phi_0^i=0$ for any $i>0$, the proof is similar to the proof of Lemma 1 or the linearized case. Finally, according to Lemma 2, the gravitational field and the Maxwell field are of type N and $l$ is the principle null direction.

Now we consider that $l$ is the principle null direction as initial data and derive the general solution space of the NP system in the NU gauge. Correspondingly, we should set
\be
\Psi_0=0,\quad \Psi_1=0,\quad \Psi_2=0,\quad \Psi_3=0. 
\ee
The solution space is given by
\begin{equation}\label{typensolution}
\begin{split}
    &\sigma=0,\quad \rho=-\frac{1}{r},\quad L^{\bz}=\frac{P}{r},\quad \alpha=\frac{\alpha^0}{r},\quad \alpha^0=\frac12 \bP \p \ln P,\quad \beta=-\xbar\alpha,\\
    &\tau=0,\quad \omega=0,\quad X^A=0,\quad \lambda=0,\quad   \mu=\frac{\mu^0}{r},\quad \mu^0= - \eth \alpha^0 - \xbar\eth\xbar\alpha^0=\eth\xbar\eth \varphi,\\ &\gamma=\gamma^0,\quad \gamma^0=-\frac12 \p_u \ln \bP,\quad \nu=\nu^0,\quad \nu^0=\xbar\eth(\gamma^0+\xbar\gamma^0)=\xbar\eth\p_u\varphi,\\
    &U=-(\gamma^0+\xbar\gamma^0)r+\mu^0,\quad \Psi_4=\frac{\Psi_4^0}{r},\quad \Psi_4^0=\xbar\eth^2\p_u\varphi,
\end{split}
\end{equation}
where we define $P\bP=2e^{-2\varphi}$.
The constraints on $\varphi$ from the NP equations are
\be
\eth^2\xbar\eth\varphi=0=\xbar\eth^2\eth\varphi,
\ee
which yields the following equation
\be
\eth\xbar\eth \varphi=2e^{-2\varphi} \p\bp \varphi=f_0(u).
\ee
Note that
\be
\eth \Psi_4^0=(\p_u + 4 \gamma^0 + 2 \xbar\gamma^0)\xbar\eth^2\eth\varphi=0,
\ee
which does not lead to a new constraint on $\varphi$. The ``$\eth$'' operator is defined by
\begin{equation}\begin{split}
&\eth \eta^{(s)}=P\bP^{-s}\bp(\bP^s \eta^{(s)})=P\bar\p \eta^{(s)} + 2 s\xbar\alpha^0 \eta^{(s)},\\
&\xbar\eth \eta^{(s)}=\bP P^{s}\p(P^{-s} \eta^{(s)})=\bP\p \eta^{(s)} -2 s \alpha^0 \eta^{(s)},\label{eth}
\end{split}\end{equation}
where $s$ is the spin weight of the field $\eta^{(s)}$. The spin weights of relevant fields are listed in Table \ref{t1}.
\begin{table}[ht]
\caption{Spin weights}\label{t1}
\begin{center}\begin{tabular}{|c|c|c|c|c|c|c|c|c|c|c|c|c|c|c|c|c|c|c|c|c}
\hline
& $\eth$ & $\p_u$ & $\alpha^0$ & $\gamma^0$ & $\nu^0$ & $\mu^0$ & $\sigma^0$ & $\lambda^0$  & $\Psi^0_4$ &  $\Psi^0_3$ & $\Psi^0_2$ & $\Psi^0_1$ & $\Psi_0^0$ & $\phi_2$ & $\phi_1$ & $\phi_0$   \\
\hline
s & $1$& $0$& $-1$ &$0$& $-1$& $0$& $2$& $-2$  &
  $-2 $&  $-1$ & $0$ & $1$ & $2$   &  $-1$ & $0$ & $1$ \\
\hline
\end{tabular}\end{center}\end{table}
The operators $\eth$,
$\overline{\eth}$ raise and lower the spin weight by one unit. Note that $P$ is of spin weight $1$ and ``holomorphic'',
$\xbar\eth P = 0$ and the commutator of the operators is
\begin{equation}
[\xbar\eth, \eth] \eta^{(s)} = \frac{s}{2} R \eta^{(s)},
\end{equation} with $R= -4\mu^0$. We also have their commutation relation with the time derivative \cite{Barnich:2019vzx},
\begin{equation}
[\partial_u, \eth] \eta^{(s)} =-2 (\xbar\gamma^0 \eth + s \eth
\gamma^0)\eta^{(s)},\quad [\p_u,\xbar\eth]\eta^{(s)}=-2(\gamma^0\xbar\eth -
s\xbar\eth\xbar\gamma^0)\eta^{(s)}.
\end{equation}
The solution \eqref{typensolution} is the type N Robinson-Trautman solution \cite{Stephani:2003tm}. The Weyl double copy of the Robinson-Trautman solution is revealed in \cite{Godazgar:2020zbv} for a special case by $\mu_0=0,\pm1$.

With the type N spacetime \eqref{typensolution} as the background and the conditions $\phi_0 = \phi_1 = 0$, it is straightforward to solve the Maxwell's equations. The solution is
\begin{equation}
\phi_2=\frac{\phi_2^0}{r},\quad \eth \phi_2^0=0, \quad \phi_2^0=\bar{P} f_\phi(u,z),
\end{equation}
where $f_\phi(u,z)$ is an arbitrary function. The WDC relation $\Psi_4=F \phi_2 \phi_2$ yields
\be
F=\frac{\Psi_4^0}{ \phi_2^0 \phi_2^0} r.
\ee
From the original WDC relations in \eqref{wdcnp}, $S=\frac{3c\phi_2^0 \phi_2^0}{\Psi_4^0 r}$. Note that $\eth \Psi_4^0=0$. This yields that $\Psi_4^0=\bP^2 f_\Psi(u,z)$. Hence, $S$ must be of the form $\frac{\zeta(u,z)}{r}$, which is a solution of the equation of motion for a scalar field on both Minkowski spacetime and the curved spacetime \eqref{typensolution}.

\subsection{Type D solution}

In this subsection, we will demonstrate for the case $\phi_2^0=\phi_0^0=0$, $\phi_1^0\neq0$. The asymptotic form of the Maxwell fields are now
\begin{equation}
\phi_0=\frac{\phi_0^1}{r^4}+O(r^{-5}),\quad \phi_1=\frac{\phi_1^0}{r^2}+O(r^{-4}),\quad \phi_2=-\frac{\bar{\eth}\phi_1^0}{r^2}+O(r^{-3}).
\end{equation}
The WDC relations lead to
\begin{equation}
\Psi_0=\frac{F_0(\phi_0^1)^2}{r^7}+O(r^{-8}), \quad \Psi_1=\frac{F_0\phi_1^0\phi_0^1}{r^5}+O(r^{-6}).
\end{equation}
The peeling-off property encoed in \eqref{R14} requires that $\phi_0^1=0$ when $\phi_1^0\neq0$. Following the same argument, any $\phi_0^i$ are thus equal to zero, therefore $\phi_0=0$. Then, the WDC relations determine that $\Psi_0=\Psi_1=0$, and the Goldberg-Sachs theorem yields that $\sigma=0$, hence $\rho=-\frac{1}{r}$. Substituting these results back into the Maxwell's equations, we obtain
\be
\phi_1=\frac{\phi_1^0}{r^2}.
\ee
We can use a first class of null rotation to turn off $\phi_2$. Then $l$ and the rotated basis $n'$ are the principle null directions and the gravitational solution is of type D in the manifest form. However, the null rotation will turn on the spin coefficient $\pi$, which are out of the NU gauge.

Nevertheless, we will still adopt the NU gauge to derive the gravitational solution space. Correspondingly, we will have
\begin{equation}
    \sigma=0,\quad\Psi_0=0,\quad \Psi_1=0, \quad \Psi_3=\cO(r^{-3}), \quad \Psi_4=\cO(r^{-3}).
\end{equation}
The solution space of gravitational theory is given by
\begin{equation}\label{typedsolution}
\begin{split}
    &\sigma=0,\quad \rho=-\frac{1}{r},\quad L^{\bz}=\frac{P}{r},\quad \alpha=\frac{\alpha^0}{r},\quad \alpha^0=\frac12 \bP \p \ln P,\quad \beta=-\xbar\alpha,\\
    &\tau=0,\quad \omega=0,\quad X^A=0,\quad \lambda=0,\quad \Psi_3=0,\quad \Psi_4=0, \\
    &\mu=\frac{\mu^0}{r}-\frac{\Psi_2^0}{r^2},\quad \mu^0= - \eth \alpha^0 - \xbar\eth\xbar\alpha^0=\eth\xbar\eth \varphi,\quad \varphi=\varphi_0(u) +\xbar\varphi(z,\bz),\\ 
    &\Psi_2=\frac{\Psi_2^0(u)}{r^3},\quad \Psi_2^0=C_\Psi e^{-3\varphi_0},\quad \xbar\Psi_2^0=\Psi_2^0,\\ 
    &\gamma=\gamma^0-\frac{\Psi_2^0}{2r^2},\quad \gamma^0=-\frac12 \p_u \ln \bP,\quad \nu=0,\\
    &U=-(\gamma^0+\xbar\gamma^0)r+\mu^0-\frac{\Psi_2^0}{r},
\end{split}
\end{equation}
where $\xbar\varphi$ is constrained by $e^{-2\xbar\varphi} \p\bp \xbar\varphi=C_\varphi$, and $C_\varphi$, $C_\Psi$ are two arbitrary constants.

With the spacetime \eqref{typedsolution} as the background and the conditions $\phi_0=0=\phi_2^0$, it
is straightforward to solve the Maxwell’s equations as
\be
\phi_1=\frac{\phi_1^0}{r^2},\quad \phi_2=-\frac{\xbar\eth \phi_1^0}{r^2},
\ee
where $\phi_1^0$ is constrained by
\be
\eth \phi_1^0=0,\quad \p_u \phi_1^0 + 2(\gamma^0+\xbar\gamma^0)\phi_1^0=0.
\ee
The WDC relations incorporated with the gravity solution \eqref{typedsolution}, yield that $\phi_2=0$. Hence, $\phi_1^0$ can be fixed as
\be
\phi_1^0=C_\phi e^{-2\varphi_0}.
\ee
Finally, the scalar field in the WDC relations is determined as 
\be
F=\frac{3e^{\varphi_0} C_\Psi}{2C_\phi^2}r.
\ee
Clearly, the corresponding scalar function $S$ satisfies the equation of motion on Minkowski spacetime. However, it is not a solution on the curved spacetime \eqref{typedsolution}. One can consider an alternative option to realize the WDC relations which involves the Maxwell theory on Minkowski spacetime \cite{Luna:2018dpt}. The solution space of the Maxwell theory for such case has been presented in Section \ref{solution}. The WDC relations yield that 
\begin{equation}
\phi_2=0=\phi_0,\quad \phi_1=\frac{C_\phi}{r^2},
\end{equation}
which leads to $\varphi_0=0$ for the gravitational solution by the WDC relations. Hence,
\be
F=\frac{3 C_\Psi}{2C_\phi^2}r,
\ee
and the corresponding scalar function $S$ satisfies the equation of motion on the flat background. 


\section{Concluding remarks}

In this paper, we study the WDC constraints on the solutions of Maxwell theory and gravity. This is an extension of the original scope of the WDC by selecting from a larger solution for the part that satisfies the WDC formula. We find that two straightforward solutions to the asymptotic WDC constraints lead to interesting truncation of the solution space. In particular, our computation shows that, when turning off the leading order of the Weyl scalar in the NP formalism in the NU gauge, WDC constraints will turn off the Weyl scalar at any order. We also present a linearized version of the WDC by checking the relations of the linearized Weyl tensor and the Maxwell field strength tensor. In this way, one may better appreciate the connection of the WDC and the Kerr-Schild double copy formula and may seek for a unified treatment for the classical double copy relation, which could be a future direction explored elsewhere.

\section*{Acknowledgments}
This work is supported in part by the National Natural Science Foundation of China (NSFC) under Grants No. 11905156 and No. 11935009.

\appendix

\section{NP formalism and NU gauge}
\label{NP}

The NP formalism \cite{Newman:1961qr} is a special tetrad formalism in four dimensions with all basis  vectors null, denoted as $e_1=l=e^2,\;e_2=n=e^1,\;e_3=m=-e^4,\;e_4=\bar{m}=-e^3$. The null basis vectors satisfy the following orthogonality and normalization conditions,
\be\label{tetradcondition}
l\cdot m=l\cdot\bm=n\cdot m=n\cdot\bm=0,\quad l\cdot n=1,\quad m\cdot\bm=-1,
\ee
where $l$ and $n$ are real, while $m$ and $\bm$ are complex conjugates of each other.
The 24 components of spin connection are interpreted as 12 complex scalars labeled by Greek symbols. The components of the Weyl tensor are represented by five complex scalars, denoted as $\Psi_0$, $\Psi_1$, $\Psi_2$, $\Psi_3$, $\Psi_4$. We would refer to \cite{Chandrasekhar} for the full notations. We will deal with vacuum solution in this work where the Ricci tensor vanishes. 

In the NP formalism, it is of great convenience to impose the UN gauge \cite{Newman:1962cia}, where one chooses
\begin{align}
\pi=\kappa=\epsilon=0,\,\,\;\;\rho=\bar\rho,\;\;\,\,\tau=\bar\alpha+\beta.
\end{align}
The geometric meaning of the NU gauge is that $l$ is tangent to a null geodesic with affine parameter and is the gradient of a scalar field. Moreover, other null basis vectors are parallel transported along $l$. Then, the coordinate system is selected as follows. The scalar field defining $l$ is chosen as coordinate $u=x^1$ and the affine parameter of the geodesic of $l$ is chosen as coordinate $r=x^2$. The rest two angular coordinates are the stereographic coordinates $A=(z,\bz)$ and they are related to the usual angular variables $(\theta,\phi)$ by $z=\cot\frac\theta2 e^{i\phi}$. In this coordinate system, the tetrad and the co-tetrad satisfying conditions in \eqref{tetradcondition} must have the forms
\begin{equation}
\begin{split}\label{tetrad}
&n=\frac{\p}{\p u} + U \frac{\p}{\p r} + X^A \frac{\p}{\p x^A},\\
&l=\frac{\p}{\p r},\\
&m=\omega\frac{\p}{\p r} + L^A \frac{\p}{\p x^A}.
\end{split}
\end{equation}
The basis vectors are assigned with special symbols as diretional derivatives,
\begin{align}
D=l^\mu\p_\mu,\;\;\;\;\Delta=n^\mu\p_\mu,\;\;\;\;\delta=m^\mu\p_\mu.
\end{align}
The NP equations are organized as follows:

\noindent
\textbf{Radial equations}
\bea
&&D\rho =\rho^2+\sigma\xbar\sigma,\label{R1}\\
&&D\sigma=2\rho \sigma + \Psi_{0},\label{R2}\\
&&D\tau =\tau \rho +  \xbar \tau \sigma   + \Psi_1 ,\label{R3}\\
&&D\alpha=\rho  \alpha + \beta \xbar \sigma  ,\label{R4}\\
&&D\beta  =\alpha \sigma + \rho  \beta + \Psi_{1},\label{R5}\\
&&D\gamma=\tau \alpha +  \xbar \tau \beta  + \Psi_2,\label{R6}\\
&&D\lambda=\rho\lambda + \xbar\sigma\mu ,\label{R7}\\
&&D\mu =\rho \mu + \sigma\lambda + \Psi_{2},\label{R8}\\
&&D\nu =\xbar\tau \mu + \tau  \lambda + \Psi_3,\label{R9}\\
&&DU=\xbar\tau\omega+\tau\xbar\omega - (\gamma+\xbar\gamma),\label{R10}\\
&&DX^A=\xbar\tau L^A + \tau\bar L^A,\label{R11}\\
&&D\omega=\rho\omega+\sigma\xbar\omega-\tau,\label{R12}\\
&&DL^A=\rho L^A + \sigma \bar L^A,\label{R13}\\
&&D\Psi_1 - \xbar\delta \Psi_0 =  4 \rho \Psi_1 - 4\alpha \Psi_0,\label{R14}\\
&&D\Psi_2 - \xbar\delta \Psi_1 =   3\rho \Psi_2  - 2 \alpha \Psi_1- \lambda \Psi_0,\label{R15}\\
&&D\Psi_3 - \xbar\delta \Psi_2 =  2\rho \Psi_3 - 2\lambda \Psi_1,\label{R16}\\
&&D\Psi_4 - \xbar\delta \Psi_3 = \rho  \Psi_4 + 2 \alpha \Psi_3 - 3 \lambda \Psi_2,\label{R17}
\eea
\textbf{Non-radial equations}
\bea
&&\Delta\lambda  = \xbar\delta\nu- (\mu + \xbar\mu)\lambda - (3\gamma - \xbar\gamma)\lambda + 2\alpha \nu - \Psi_4,\label{H1}\\
&&\Delta\rho= \xbar\delta\tau- \rho\xbar\mu - \sigma\lambda  -2\alpha \tau + (\gamma + \xbar\gamma)\rho  - \Psi_2,\label{H2}\\
&&\Delta\alpha = \xbar\delta\gamma +\rho \nu - (\tau + \beta)\lambda + (\xbar\gamma - \gamma -\xbar \mu)\alpha  -\Psi_3 ,\label{H3}\\
&&\Delta \mu=\delta\nu-\mu^2 - \lambda\xbar\lambda - (\gamma + \xbar\gamma)\mu   + 2 \beta \nu ,\label{H4}\\
&&\Delta \beta=\delta\gamma - \mu\tau + \sigma\nu + \beta(\gamma - \xbar\gamma -\mu) - \alpha\xbar\lambda ,\label{H5}\\
&&\Delta \sigma=\delta\tau - \sigma\mu - \rho\xbar\lambda - 2 \beta \tau + (3\gamma - \xbar\gamma)\sigma  ,\label{H6}\\
&&\Delta \omega=\delta U +\xbar\nu -\xbar\lambda\xbar\omega + (\gamma-\xbar\gamma-\mu)\omega,\label{H7}\\
&&\Delta L^A=\delta X^A - \xbar\lambda \bar L^A + (\gamma-\xbar\gamma-\mu)L^A,\label{H8}\\
&&\delta\rho - \xbar\delta\sigma=\rho\tau - \sigma (3\alpha - \xbar\beta)   - \Psi_1 ,\label{H9}\\
&&\delta\alpha - \xbar\delta\beta=\mu\rho - \lambda\sigma + \alpha\xbar\alpha + \beta\xbar\beta - 2 \alpha\beta - \Psi_2,\label{H10}\\
&&\delta\lambda - \xbar\delta\mu= \mu \xbar\tau + \lambda (\xbar\alpha - 3\beta) - \Psi_3 ,\label{H11}\\
&&\delta \xbar\omega-\bar\delta\omega=\mu - \xbar\mu - (\alpha - \xbar\beta) \omega +  (\xbar\alpha - \beta)\xbar\omega,\label{H12}\\
&&\delta \bar L^A - \bar\delta L^A= (\xbar\alpha - \beta)\bar L^A -  (\alpha - \xbar\beta) L^A ,\label{H13}\\
&&\Delta\Psi_0 - \delta \Psi_1 = (4\gamma -\mu)\Psi_0 - (4\tau + 2\beta)\Psi_1 + 3\sigma \Psi_2,\label{H14}\\
&&\Delta\Psi_1 - \delta \Psi_2 = \nu\Psi_0 + (2\gamma - 2\mu)\Psi_1 - 3\tau \Psi_2 + 2\sigma \Psi_3 ,\label{H15}\\
&&\Delta\Psi_2 - \delta \Psi_3 = 2\nu \Psi_1 - 3\mu \Psi_2 + (2\beta - 2\tau) \Psi_3 + \sigma \Psi_4,\label{H16}\\
&&\Delta\Psi_3 - \delta \Psi_4 = 3\nu \Psi_2 - (2\gamma + 4\mu) \Psi_3 + (4\beta - \tau) \Psi_4,\label{H17}
\eea
\textbf{Maxwell's equations}
\bea
&&D\phi_1 - \xbar\delta \phi_0 = 2\rho \phi_1 - 2\alpha \phi_0,\label{R18}\\
&&D\phi_2 - \xbar\delta \phi_1 = \rho \phi_2 - \lambda \phi_0,\label{R19}\\
&&\Delta\phi_0 - \delta \phi_1 = (2\gamma-\mu) \phi_0 - 2\tau \phi_1 + \sigma \phi_2,\label{H18}\\
&&\Delta\phi_1 - \delta \phi_2 = \nu \phi_0 - 2 \mu \phi_1 - (\xbar\alpha - \beta) \phi_2.\label{H19}
\eea

\bibliography{ref}
\end{document}